\documentclass[a4paper,10pt,nofootinbib]{revtex4}
\usepackage{graphicx}  
\usepackage{amsmath}   
\usepackage{amssymb}   
\usepackage{bm} 
\usepackage{dcolumn}
\usepackage{color}
\usepackage{mathrsfs}
\usepackage{amsfonts}
\usepackage{varioref}
\usepackage{mathrsfs}
\usepackage{graphicx}
\usepackage{latexsym}
\usepackage{amsmath}
\usepackage{amssymb}
\usepackage{textcomp}
\usepackage{amsbsy}
\usepackage{graphics}
\usepackage{epstopdf}
\usepackage{color}
\usepackage[caption=false]{subfig}
\usepackage{enumitem}
\usepackage{array,multirow} 

\usepackage{float}

\RequirePackage[colorlinks,citecolor=blue,urlcolor=magenta,linkcolor=blue,hyperfootnotes=true]{hyperref}
\input epsf

\allowdisplaybreaks[4]

\begin{document}

\tolerance=5000

\title{Different aspects of entropic cosmology}

\author{Shin'ichi~Nojiri$^{1,2}$\,\thanks{nojiri@gravity.phys.nagoya-u.ac.jp},
Sergei~D.~Odintsov$^{3,4}$\,\thanks{odintsov@ieec.cat},
Tanmoy~Paul$^{5}$\,\thanks{pul.tnmy9@gmail.com}} \affiliation{
$^{1)}$ KEK Theory Center, Institute of Particle and Nuclear Studies,
High Energy Accelerator Research Organization (KEK), Oho 1-1, Tsukuba, Ibaraki 305-0801, Japan \\
$^{2)}$ Kobayashi-Maskawa Institute for the Origin of Particles
and the Universe, Nagoya University, Nagoya 464-8602, Japan \\
$^{3)}$ ICREA, Passeig Lluis Companys, 23, 08010 Barcelona, Spain\\
$^{4)}$ Institute of Space Sciences (ICE, CSIC) C. Can Magrans s/n, 08193 Barcelona, Spain\\
$^{5)}$ Department of Physics, Visva-Bharati University, Santiniketan 731235, India}

\begin{abstract}
We give a short review of the recent developments of entropic cosmology based on two thermodynamic laws of the apparent horizon, namely the first and the second laws of thermodynamics. The first law essentially provides the change of the entropy of the apparent horizon during the cosmic evolution of the universe, in particular, it is given by: $TdS = -d(\rho V) + W dV$ (where $W$ is the work density 
and other quantities have their usual meaning). In this way, the first law actually links various theories of gravity with the entropy of the apparent horizon. This leads to a natural question --- ``what will be the form of the horizon entropy corresponding to a general modified theory of gravity?'' The second law of horizon thermodynamics states that the change of total entropy (the sum of horizon entropy $+$ matter fields' entropy) with respect to cosmic time must be positive, where the matter fields behave like an open system characterised by a non-zero chemical potential. The second law of horizon thermodynamics importantly provides model-independent constraints on entropic parameters. Finally, we discuss the standpoint of entropic cosmology on inflation (or bounce), reheating and primordial gravitational waves from the perspective of generalised entropy function.
\end{abstract}

\maketitle

\section{Introduction}

The benchmark of the Bekenstein-Hawking entropy is that it connects two apparently different sectors, namely, gravity and thermodynamics, on equal footing. In particular, the black hole horizon has a thermal behaviour where the entropy of the horizon scales by the area of the horizon, and the surface gravity fixes the corresponding temperature \cite{Bekenstein:1973ur, Hawking:1975vcx, Bardeen:1973gs, Wald:1999vt}. In the cosmological context, there exists an apparent horizon which is a marginally trapped surface with vanishing expansion and divides the observable universe from the unobservable one. Similar to black hole thermodynamics, the cosmic horizon is generally considered to have a thermal behaviour \cite{Cai:2005ra, Akbar:2006kj, Cai:2006rs, Akbar:2006er,
Paranjape:2006ca, Jamil:2009eb, Cai:2009ph, Wang:2009zv, Jamil:2010di, Gim:2014nba, DAgostino:2019wko, Sanchez:2022xfh, Cognola:2005de, Nojiri:2022nmu, Nojiri:2023wzz, Odintsov:2024hzu, Jizba:2023fkp, Jizba:2024klq, Nojiri:2022aof, Nojiri:2022dkr, Odintsov:2022qnn}, and one can motivate it by the following arguments:
\begin{itemize}
 \item During the cosmic evolution of the universe, the matter fields inside of the horizon show a flux from inside to outside the horizon (the flux is outward in nature during the accelerating stage), which results in a decrease in the matter fields' entropy. This violates the second law of thermodynamics which states that the change of total entropy must be positive. Therefore the cosmic horizon should be incorporated with some entropy to manage the increase of total entropy (the sum of horizon entropy + matter fields' entropy).

 \item The Cosmological field equations are time reversal symmetric and thus they always come with a contracting solution along with an expanding one. However, our observational data indicate that the universe is expanding. Therefore the natural question that comes to mind --- ``Why does the universe always choose the expanding solution?'' In order to answer this question, we need to associate thermal behaviour with the cosmic horizon. Then the second law of horizon thermodynamics actually disagrees the contracting solution in order to have a positive change of the total entropy.
\end{itemize}
Consequently the subject ``entropic cosmology'' gained a lot of interest, where the cosmic horizon is associated with an entropy that follows the thermodynamic law \cite{Cai:2005ra, Akbar:2006kj, Cai:2006rs, Akbar:2006er, Nojiri:2023wzz}:
\begin{eqnarray}
T_\mathrm{h}dS_\mathrm{h} = -d\left(\rho V\right) + \frac{1}{2}\left(\rho - p\right)dV
\label{law-1}
\end{eqnarray}
where $V$ is the volume enclosed by the apparent horizon given by $R_\mathrm{h} = 1/H$ (with $H$ being the Hubble parameter of the universe), moreover $T_\mathrm{h}$ and $S_\mathrm{h}$ are the temperature and the entropy of the horizon, respectively (and the other quantities have their usual meaning). We now assume that the universe is homogeneous and isotropic, owing to which, the total energy inside the apparent horizon is expressed as $U =\rho V$, otherwise, if $\rho$ depends on spatial coordinates then the energy inside the horizon should be expressed by the integral (over the volume of the horizon) $U = \int \rho dV$. Here $T_\mathrm{h}$ is fixed by the surface gravity of the apparent horizon, which, in the case of spatially flat Friedmann-Lema\^{i}tre-Robertson-Walker (FLRW) metric, turns out to be,
\begin{eqnarray}
 T_\mathrm{h} = \frac{H}{2\pi}\left|1 + \frac{\dot{H}}{2H^2}\right|~~.
 \label{AH2}
\end{eqnarray}
At this stage, we would like to mention that the microscopic origin of horizon entropy (and its associated temperature) is still a debatable topic and needs further investigation (one may see \cite{Nojiri:2023bom, Housset:2023jcm} for some progress in this regard). Together with Eq.~(\ref{law-1}) (which is the first law of horizon thermodynamics in cosmology), the second law states that \cite{Odintsov:2024hzu}
\begin{eqnarray}
 d\left(S_\mathrm{h} + S_\mathrm{m}\right) > 0~,
 \label{law-2}
\end{eqnarray}
where $S_\mathrm{m}$ represents the entropy of the matter fields inside of the horizon. With a specific form of horizon entropy, the first law of horizon thermodynamics (along with the local conservation law of the matter fields) leads to the cosmological field equation; for instance, the Bekenstein-Hawking like the entropy of the cosmic horizon provides the usual FLRW equations of Einstein gravity from Eq.~(\ref{law-1}). However for a different form of horizon entropy (compared to the Bekenstein-Hawking like one), then we land up with modified cosmological field equations:
\begin{eqnarray}
\dot{H}\left(\frac{\partial S_\mathrm{h}}{\partial S}\right) = -4\pi G\left(\rho + p\right)
\label{HE-4}
\end{eqnarray}
and
\begin{eqnarray}
 \int \left(\frac{\partial S_\mathrm{h}}{\partial S}\right) d\left(H^2\right) = \frac{8\pi G}{3}\rho + \frac{\Lambda}{3}~~,
 \label{HE-5}
\end{eqnarray}
respectively, where $S = \pi/(GH^2)$ is the Bekenstein-Hawking entropy (clearly for $S_\mathrm{h} \equiv S$, one obtains the usual FLRW equations). Such modified cosmic scenario has some interesting cosmological consequences started from inflation (or bounce) to dark energy era \cite{Cai:2005ra, Akbar:2006kj, Cai:2006rs, Akbar:2006er, Paranjape:2006ca, Jamil:2009eb, Cai:2009ph, Wang:2009zv, Jamil:2010di, Gim:2014nba, DAgostino:2019wko, Sanchez:2022xfh, Cognola:2005de, Nojiri:2022nmu, Nojiri:2023wzz, Odintsov:2024hzu, Jizba:2023fkp, Jizba:2024klq, Nojiri:2022aof, Nojiri:2022dkr, Odintsov:2022qnn, Odintsov:2023vpj, Bolotin:2023wiw, Lymperis:2023prf, Volovik:2024eni, Volovik:2023wty, Odintsov:2024sbo, Brevik:2024nzf, Okcu:2024tnw, Cruz:2023xjp, Cardenas:2023zmn, Luciano:2024bco, DAgostino:2024sgm}. Some variants of the Bekenstein-Hawking entropy are the Tsallis entropy \cite{Tsallis:1987eu}, the R\'{e}nyi entropy \cite{Renyi}, the Barrow entropy \cite{Barrow:2020tzx}, the Sharma-Mittal entropy \cite{SayahianJahromi:2018irq}, the Kaniadakis entropy \cite{Kaniadakis:2005zk}, the Loop Quantum Gravity entropy \cite{Liu:2021dvj}, etc. The important point is that all of these entropies have some common features like --- they all vanish at the limit $S \rightarrow 0$ and they show a monotonic increasing behaviour with respect to the Bekenstein-Hawking entropy variable ($S$). One more interest in entropic cosmology is that it is related to holographic cosmology initiated by Witten and Susskind \cite{Witten:1998qj, Susskind:1998dq, Fischler:1998st}. In particular, entropic cosmology proves to be equivalent to the generalised holographic scenario with suitable holographic cut-offs \cite{Nojiri:2021iko, Nojiri:2021jxf}. The significant contributions of holographic cosmology (or equivalently, entropic cosmology) corresponding to the aforementioned entropies is the explanation of the dark energy era of our universe, namely the holographic dark energy (HDE) \cite{Li:2004rb, Li:2011sd, Wang:2016och, Nojiri:2005pu, Enqvist:2004xv, Gong:2004cb, Gao:2007ep, Li:2009bn, Zhang:2009un, Lu:2009iv, Komatsu:2016vof, Komatsu:2015nkb, Nojiri:2019skr, Barrow:2020kug, Nojiri:2023nop, Nojiri:2020wmh, Bhardwaj:2021chg, Chakraborty:2020jsq, Sarkar:2021izd}.\\

Based on the above arguments, some immediate questions that arise are as follows:
\begin{enumerate}
 \item What will be the form of the horizon entropy that leads to the cosmological field equations for a general modified theory of gravity from Eq.~(\ref{law-1})?

 \item Does there exist any generalised entropy that can generalise all the known entropies proposed so far (like the Tsallis entropy, the R\'{e}nyi entropy, the Barrow entropy, the Sharma-Mittal entropy, the Kaniadakis entropy etc.)? This question is well motivated as all these entropies share some common properties mentioned above.

 \item If a generalised form of entropy exists, then what are the constraints on the generalised entropic parameters coming from the second law of horizon thermodynamics? Furthermore, what is the standpoint of generalised entropy on primordial gravitational waves? Does the constraint coming from the primordial gravitational waves match with that of based on the second law of horizon thermodynamics?
\end{enumerate}
The present article, based on some of our previous works \cite{Nojiri:2022dkr, Odintsov:2022qnn, Odintsov:2023vpj, Nojiri:2023wzz, Odintsov:2024hzu, Odintsov:2024sbo}, gives a brief review in answering the above questions. We will follow the $\left(-,+,+,+,....\right)$ signature of a spatially flat $(n+1)$ dimensional spacetime metric, and we take $G = c= \hbar =1$ where $G$ is the Newton's constant, $c$ is the speed of light and $\hbar$ is Planck constant.

\section{First law of horizon thermodynamics: Consistent entropy for a general modified theory of gravity}\label{sec-law-1-MGT}

The question that we will encounter in this section is the following: what is the form of entropy which, based on the thermodynamic law (\ref{law-1}),
can produce the cosmological field equations for a general modified theory of gravity \cite{Nojiri:2023wzz}?

The FLRW equations for a general modified theory of gravity in $(n+1)$ dimensional spacetime can be expressed as,
\begin{align}
H^2=&\, \frac{16\pi}{n(n-1)}(\rho + \rho_\mathrm{c})~~~~~~\mathrm{and}~~~~~~\dot{H}=-\frac{8\pi}{(n-1)}(\rho + \rho_\mathrm{c} + p + p_\mathrm{c})\, ,
\label{FRW-MGT}
\end{align}
where $\rho_\mathrm{c}$ and $p_\mathrm{c}$ represent the modifications compared to the Einstein gravity. Owing to such modifications, we may expect that the corresponding horizon entropy for a general modified gravity theory will take the following form,
\begin{align}
S_\mathrm{h} = \frac{A}{4} + S_\mathrm{c}(A)\, ,
\label{correct entropy-MGT-1}
\end{align}
with the suffix `h' for `horizon' entropy. Here $A = n\Omega_n{R_\mathrm{h}}^{n-1}$ represents the area of the apparent horizon in $n+1$ dimensional spacetime, and $S_\mathrm{c}$ is a function of $A$. 
In particular, the horizon entropy for general gravity theory is considered to be corrected over that of in the case of Einstein gravity, namely Eq.~(\ref{correct entropy-MGT-1}). The correction $S_\mathrm{c}$ \textrm{explicitly} depends on the modification of gravitational action, which we need to find in such a way that the following thermodynamic law, namely
\begin{align}
T_\mathrm{h}dS_\mathrm{h} = -dE + WdV\, ,
\label{MGT-1}
\end{align}
holds, with $E = \rho V$ and $W = \frac{1}{2}\left(\rho - p\right)$.
By using $S_\mathrm{h} = \frac{A}{4} + S_\mathrm{c}$, the above equation can be equivalently written as,
\begin{align}
T_\mathrm{h}\frac{dS_\mathrm{c}}{dt} = \frac{d}{dt}\left(\rho_\mathrm{c}V\right) - W_\mathrm{c}\frac{dV}{dt}
+ \left[-\frac{d}{dt}\left(\rho V + \rho_\mathrm{c}V\right) + \frac{1}{2}\left(\rho + \rho_\mathrm{c} - p - p_\mathrm{c}\right)\frac{dV}{dt} - T\frac{d}{dt}\left(\frac{A}{4}\right)\right]\, ,
\label{MGT-2}
\end{align}
where $W_\mathrm{c} = \frac{1}{2}\left(\rho_\mathrm{c} - p_\mathrm{c}\right)$. With $R_\mathrm{h} = \frac{1}{H}$ along with Eq.~(\ref{AH2}), we obtain
\begin{align}
\frac{dS_\mathrm{c}}{dt} = -2\pi n\Omega_n\left(\rho_\mathrm{c} + p_\mathrm{c}\right){R_\mathrm{h}}^n\, ,
\label{MGT-6}
\end{align}
on integrating which, we obtain, 
\begin{align}
S_\mathrm{c} = 2\pi n\Omega_n\int {R_\mathrm{h}}^{n-2}\left(\frac{\rho_\mathrm{c} + p_\mathrm{c}}{\dot{H}}\right)dR_\mathrm{h}\, .
\label{MGT-7}
\end{align}
Eq.~(\ref{MGT-7}) argues that $S_\mathrm{c}$ should be the function of only $R_\mathrm{h}$ (or equivalently, the area of the horizon) as $dR_\mathrm{h}$ is the only differential present in the rhs of the above expression. In general, the integration in Eq.~(\ref{MGT-7}) should be realized by specifying the scale factor $a = a(t)$ as a function of the cosmological time $t$. In particular, if we consider a specific scale factor, then $H(t)$ and consequently $R_\mathrm{h}$ are also given by function of $t$, and as a result, the integrand in Eq.~(\ref{MGT-7}) may be expressed in terms of $R_\mathrm{h}$. Then Eq.~(\ref{MGT-7}) can be integrated to give $S_\mathrm{c} = S_\mathrm{c}(R_\mathrm{h})$.
With the above form of $S_\mathrm{c}$, Eq.~(\ref{correct entropy-MGT-1}) provides the full entropy corresponding to a general modified gravity theory as,
\begin{align}
S_\mathrm{h} = \frac{A}{4} + 2\pi n\Omega_n\int {R_\mathrm{h}}^{n-2}\left(\frac{\rho_\mathrm{c} + p_\mathrm{c}}{\dot{H}}\right)dR_\mathrm{h}\, ,
\label{MGT-8}
\end{align}
It may be observed that for $\rho_\mathrm{c} = p_\mathrm{c} = 0$, i.e., for Einstein's gravity theory, the entropy from Eq.~(\ref{MGT-8}) becomes $S_\mathrm{h} = A/4$, as per our expectation. Moreover for $\rho = p = 0$, i.e., without any matter fields, Eq.~(\ref{FRW-MGT}) gives $\left(\rho_\mathrm{c} + p_\mathrm{c}\right)/\dot{H} = -8\pi/(n-1)$ which immediately yields to $S_\mathrm{h} = 0$ from Eq.~(\ref{MGT-8}). This is, however, expected as there is no flux of matter fields from inside to outside of the horizon for $\rho = p = 0$, or equivalently
there is no information loss associated with the horizon.

Below we will present some specific examples of gravity theories and will determine the respective entropy from Eq.~(\ref{MGT-8}).
\begin{itemize}
 \item For $(n+1)$ dimensional GB gravity where the FLRW equations are given by,
\begin{eqnarray}
H^2 + \lambda (n-2)(n-3) H^4&=&\frac{16\pi}{n(n-1)}\rho\, ,\nonumber\\
\left[1 + 2\lambda (n-2)(n-3) H^2\right]\dot{H}&=&-\frac{8\pi}{(n-1)}\left(\rho + p\right)\, ,
\label{FRW-GB}
\end{eqnarray}
with $\lambda$ being the GB parameter, the corresponding horizon entropy from Eq.~(\ref{MGT-8}) comes as,
\begin{align}
S_\mathrm{h} =  \frac{A}{4}\left\{1 + \frac{2\lambda (n-1)(n-2)}{{R_\mathrm{h}}^2}\right\}\, .
\label{II-4}
\end{align}

\item For $(3+1)$ dimensional $f(Q)$ gravity theory, the FLRW equations are given by,
\begin{eqnarray}
 H^2&=&\frac{1}{6f_\mathrm{Q}}\left(\rho + \frac{f}{2}\right)~,\nonumber\\
 \dot{H}&=&-\frac{1}{2f_\mathrm{Q}}\left(\frac{\rho}{2} + \frac{f}{2} + 12H^2f_\mathrm{QQ}\dot{H} - \frac{f}{4} + \frac{p}{2} - \dot{H}f_\mathrm{Q}\right)~~,
 \label{FRW-fQ}
\end{eqnarray}
where $f_\mathrm{Q}$ and $f_\mathrm{QQ}$ represent the first and second derivative of $f(Q)$ (with respect to the variable $Q$) respectively. Clearly in this case, one requires a certain form of $f(Q)$ which is taken to be a power law type: $f(Q) = Q^{n}$ (with $n$ being a constant). Such form of $f(Q)$ along with Eq.~(\ref{MGT-8}) leads to the following entropy corresponding to the $(3+1)$ dimensional $f(Q)$ gravity:
\begin{eqnarray}
 S_\mathrm{h} = \frac{A}{4}\left[1 - 32\pi\left\{1 + \frac{n(1-2n)}{(2-n)}\left(\frac{6\pi}{A}\right)^{n-1}\right\}\right]~~.
 \label{fQ-6}
\end{eqnarray}
\end{itemize}
It may be noted that in both GB gravity and $f(Q)$ gravity theories, the integration of Eq.~(\ref{MGT-8}) can be performed without specifying the scale factor $a = a(t)$. This may not be the case when the gravitational field equations contain higher derivatives of Hubble parameter, for instance --- the $F(R)$ gravity theory where the FLRW equations contain $\dddot{H}$ and thus one needs to specify the scale factor $a(t)$ to perform the integration and to determine $S_\mathrm{h}$ (corresponding to $F(R)$ gravity) from Eq.~(\ref{MGT-8}) \cite{Nojiri:2023wzz}. On other hand, it is well known that a $F(R)$ theory can be recast to a scalar-tensor theory by a conformal transformation of the spacetime metric, where the scalar potential depends on the form of $F(R)$ under consideration. However it has been pointed in \cite{Nojiri:2023wzz} that such mathematical equivalence between $F(R)$ and scalar-tensor theory gets spoiled from the perspective of the entropy of the apparent horizon.

At this stage it deserves mentioning that the determination of horizon entropy from the thermodynamic law (\ref{law-1}) encountered a problem, in particular, it requires the quantity $\left(1 + \frac{\dot{H}}{2H^2}\right)$ to be positive, otherwise, the entropy (particularly for Einstein gravity) turns out to be negative which is impossible. The condition, namely $1 + \frac{\dot{H}}{2H^2} < 0$, may occur during reheating process where the EoS parameter of the matter field ($\omega = p/\rho$) is larger than $1/3$. Then one may argue that for such reheating era where $\omega > 1/3$, there exists no such entropy (of the horizon) that connects the FLRW equations~(\ref{FRW-MGT}) with the thermodynamic law (\ref{law-1}). In order to resolve this issue, a modified thermodynamic law has been proposed in the context of cosmology \cite{Nojiri:2023wzz}, as follows:
\begin{align}
T_\mathrm{h}dS_\mathrm{h}^{(m)} = -dE + \rho dV\, ,
\label{prob-4}
\end{align}
where $T_\mathrm{h}$ is shown in Eq.~(\ref{AH2}) and the superfix `$m$' stands for the `modified' thermodynamic law. Such modified thermodynamic law indeed resolves the aforementioned problem \cite{Nojiri:2023wzz}, and thus is considered to be more general compared to the previous one (\ref{law-1}) which, however, is a limiting case of the modified thermodynamics for $p = -\rho$.\\

This modified thermodynamic law surely affects the horizon entropy compared to that of the previous one. For instance,
\begin{itemize}
 \item In the case of $(n+1)$ dimensional Einstein gravity, the modified thermodynamic law (\ref{prob-4}) leads to the corresponding horizon entropy as,
 \begin{align}
S_\mathrm{h}^{(m)} = \frac{n(n-1)}{4}\Omega_n\int \frac{{R_\mathrm{h}}^{n-2}}{\left|1 + \frac{\dot{H}}{2H^2}\right|}dR_\mathrm{h}\, ,
\label{E-L-3}
\end{align}
which for a constant EoS parameter for the matter field, i.e., for a constant $\omega = p/\rho$, results in the following form
\begin{align}
S_\mathrm{h}^{(m)}(\mathrm{constant~\omega}) = \frac{A}{\left|4 - n - n\omega\right|}\, .
\label{E-L-4}
\end{align}
Eq.~(\ref{E-L-4}) clearly depicts that the $S_\mathrm{h}^{(m)}$ explicitly depends on the value of $\omega$. Thereby in this modified thermodynamic law, the form of the entropy corresponding to Einstein's gravity changes with the evolution era of the universe, for instance (this is unlike to the previous case where the horizon entropy for Einstein's gravity is given by $A/4$ which does not change, by its form, with the evolution of the universe):
\begin{align}
\begin{array}{ll}
S_\mathrm{h}^{(m)}(\mathrm{constant~\omega})= \frac{A}{4}\, , &
\mbox{during inflation when}\ \omega =\, -1\, ,\\
S_\mathrm{h}^{(m)}(\mathrm{constant~\omega})= \frac{A}{\left|4-n\right|}\, ,&
\mbox{during matter-dominated era when}\ \omega = 0\, ,\\
S_\mathrm{h}^{(m)}(\mathrm{constant~\omega})= \frac{A}{4\left|1-n/3\right|}\, ,&
\mbox{during radiation era when}\ \omega = 1/3\, ,
\end{array}
\nonumber
\end{align}

\item For $(n+1)$ dimensional GB gravity theory, the required entropy corresponding to (\ref{prob-4}) comes as,
\begin{align}
S_\mathrm{h}^{(m)} = \frac{A}{4}\left\{1 + \frac{2\alpha}{{R_\mathrm{h}}^2}\left(\frac{n-1}{n-3}\right)\right\}\, ,
\label{II-L-4}
\end{align}
for $H = \mathrm{constant}$, and
\begin{align}
S_\mathrm{h}^{(m)} = \left(\frac{1}{\left|1 - 1/(2h_0)\right|}\right)\frac{A}{4}\left\{1 + \frac{2\alpha}{{R_\mathrm{h}}^2}\left(\frac{n-1}{n-3}\right)\right\}\, ,
\label{II-L-5}
\end{align}
for $H = h_0/t$, with $A = n\Omega_n{R_\mathrm{h}}^{n-1}$ is the area of the apparent horizon.

\item For a general modified gravity theory, the corresponding horizon entropy coming from the modified thermodynamic law (\ref{prob-4}) is obtained as,
\begin{align}
S_\mathrm{h}^{(m)} = \frac{n(n-1)\Omega_n}{4}\int \frac{{R_\mathrm{h}}^{n-2}}{\left|1 + \frac{\dot{H}}{2H^2}\right|}dR_\mathrm{h}
+ 2\pi n\Omega_n\int {R_\mathrm{h}}^{n-2}\left(\frac{\rho_\mathrm{c} + p_\mathrm{c}}{\dot{H}\left|1 + \frac{\dot{H}}{2H^2}\right|}\right)dR_\mathrm{h}\, ,
\label{MGT-L-7}
\end{align}
The above expressions of horizon entropy (for different gravity theories) arising from the modified thermodynamic law (\ref{prob-4}) prove to exist irrespective of whether $\left(1 + \frac{\dot{H}}{2H^2}\right)$ is positive or negative.
\end{itemize}

\section{Second law of horizon thermodynamics}

Till now, we have used only the first law of thermodynamics of the apparent horizon. However, in the context of horizon thermodynamics, a consistent cosmology also demands the validity of the second law of thermodynamics, i.e., the change of total entropy (which is the sum of the horizon entropy and the entropy of the matter fields) with cosmic time should be positive \cite{Odintsov:2024hzu}:
\begin{eqnarray}
 \dot{S}_\mathrm{h} + \dot S_\mathrm{m} > 0~.
 \label{N-0}
\end{eqnarray}
Eq.~(\ref{law-1}) immediately gives the change of horizon entropy as,
\begin{eqnarray}
 \dot{S}_\mathrm{h} = \frac{8\pi}{H^3}\left(\rho + p\right)~.
 \label{HE-3}
\end{eqnarray}
Besides the thermodynamics of the apparent horizon governed by Eq.~(\ref{law-1}), we also need to consider the thermodynamics of the matter fields, in particular, the matter fields inside the apparent horizon obey the following thermodynamic law:
\begin{eqnarray}
T_\mathrm{m}dS_\mathrm{m} = d\left(\rho V\right) + pdV - \mu dN~~,
\label{ME-1}
\end{eqnarray}
where $T_\mathrm{m}$ and $S_\mathrm{m}$ represent the temperature and the entropy of the matter fields respectively; note that $T_\mathrm{m}$, in general, is different than the horizon temperature. The matter fields exhibit a flux through the horizon, which is either outward or inward depending on the background cosmic era of the universe. Owing to this flux, the matter fields behave like an open system where $\mu$ (in Eq.~(\ref{ME-1})) is the chemical potential and $dN$ represents the change in particle number (within time $dt$) inside of the horizon. Due to $V = \frac{4\pi}{3H^3}$, the above expression takes the following form,
\begin{eqnarray}
 T_\mathrm{m}\dot{S}_\mathrm{m} = -\frac{4\pi}{H^2}\left(\rho + p\right)\left\{1 + \frac{\dot{H}}{H^2}\right\} - \mu\dot{N}~~.
 \label{ME-3}
\end{eqnarray}
For the purpose of $\dot{N}$ we need to understand that the speed of the formation of apparent horizon is different than the comoving expansion speed of the universe. Actually the speed of the formation of the apparent horizon turns out to be $v_\mathrm{h} = -\dot{H}/H^2$, while the comoving speed of the universe at a physical distance $d$ from an observer is given by: $v_\mathrm{c} = Hd$. Therefore $v_\mathrm{c} = 1$ at the apparent horizon (i.e., at $d = 1/H$). Therefore $v_\mathrm{c} > v_\mathrm{h}$ when the universe undergoes through an accelerating era, while for a decelerating era, we have $v_\mathrm{c} < v_\mathrm{h}$. Hence we calculate (with $\epsilon = -\dot{H}/H^2$),
\begin{eqnarray}
 V_\mathrm{c}(t+dt) - V(t+dt)=\frac{4\pi}{3}\left(\frac{1}{H} - \frac{\dot{H}}{H^2}dt\right)^3 - \frac{4\pi}{3}\left(\frac{1}{H} + dt\right)^3 = \frac{4\pi}{H^2}\left(1 - \epsilon\right)dt~~,\label{N-1}
\end{eqnarray}
which represents the gap between the comoving volume and the volume enclosed by the horizon. Therefore we may write,
\begin{eqnarray}
 \frac{dN}{dt} = -\frac{\rho}{u}\frac{d}{dt}\left[V_\mathrm{c}(t+dt) - V(t+dt)\right]~~,
 \label{ME-5}
\end{eqnarray}
where $u$ is the energy per particle. The above expression along with Eq.~(\ref{N-1}) immediately results to
\begin{eqnarray}
 \mu \dot{N} = -\frac{4\pi \rho}{H^2}\left(1-\epsilon\right)~~.
 \label{ME-6}
\end{eqnarray}
To arrive at Eq.~(\ref{ME-6}), we have used $\mu \equiv \frac{\partial}{\partial N}\left(\mathrm{total~energy}\right) = u$. Plugging this into Eq.~(\ref{ME-3}) yields,
\begin{eqnarray}
 T_\mathrm{m}\dot{S}_\mathrm{m} = -\frac{4\pi}{H^2}\left(\rho + p\right)\left\{1 + \frac{\dot{H}}{H^2}\right\} + \frac{4\pi \rho}{H^2}\left(1-\epsilon\right)~~.
 \label{ME-7}
\end{eqnarray}

Eq.~(\ref{HE-3}) and Eq.~(\ref{ME-7}) provide the change of horizon entropy as well the change of matter fields' entropy (with respect to the cosmic time). These immediately determine the change of total entropy as,
\begin{eqnarray}
 T_\mathrm{h}\frac{dS_\mathrm{h}}{dt} + T_\mathrm{m}\frac{dS_\mathrm{m}}{dt} = -2\pi\left(\rho + p\right)\left(\frac{\dot{H}}{H^4}\right) + \frac{4\pi \rho}{H^2}\left(1-\epsilon\right)~~,
 \label{TE-2}
\end{eqnarray}
We may eliminate $\rho$ and $p$ from the above expression by using the Friedmann Eqs.~(\ref{HE-4}) and (\ref{HE-5}). As a result, we obtain,
\begin{eqnarray}
 T_\mathrm{h}\frac{dS_\mathrm{h}}{dt} + T_\mathrm{m}\frac{dS_\mathrm{m}}{dt} = \frac{\epsilon^2}{2G}\left(\frac{\partial S_\mathrm{h}}{\partial S}\right)
 - \frac{3(\epsilon - 1)}{2G}\frac{1}{H^2}~\int \left(\frac{\partial S_\mathrm{h}}{\partial S}\right) d\left(H^2\right)~~.
 \label{TE-3}
\end{eqnarray}
Eq.~(\ref{TE-3}) indicates that the total change of entropy depends on two factors: (a) the background cosmic evolution of the universe (through the Hubble parameter), and (b) the form of the horizon entropy under consideration (through $S_\mathrm{h}$). Below we consider some specific forms of horizon entropy and establish the constraints on the corresponding entropic parameters in order to validate the second law of horizon thermodynamics during a wide range of cosmic eras.
\begin{itemize}
\item For Tsallis entropy $S_\mathrm{h} \equiv S_\mathrm{T} = S^{\delta}$ (where the suffix `T' stands for Tsallis entropy and $S = \frac{\pi}{GH^2}$ is the Bekenstein-Hawking entropy), the change of total entropy from Eq.~(\ref{TE-3}) comes as,
\begin{eqnarray}
 T_\mathrm{h}\left(\frac{dS_\mathrm{T}}{dt}\right) + T_\mathrm{m}\left(\frac{dS_\mathrm{m}}{dt}\right) = \left(\frac{\delta}{2G}\right)\left(\frac{\pi}{GH^2}\right)^{\delta - 1}\left\{\epsilon^2 - \frac{3\left(\epsilon - 1\right)}{\left(2 - \delta\right)}\right\}~~.
 \label{T-4}
\end{eqnarray}

\begin{enumerate}
 \item {\underline{During inflation}}: Here $\epsilon \simeq 0$ and thus in order to have in order to $T_\mathrm{h}\dot{S}_\mathrm{T} + T_\mathrm{m}\dot{S}_\mathrm{m} > 0$ from Eq.~(\ref{T-4}), the Tsallis exponent has to fulfil,
 \begin{eqnarray}
  0 < \delta < 2~~.
  \label{T-6}
 \end{eqnarray}

 \item {\underline{During reheating stage}}: Here $\epsilon = \frac{3}{2}\left(1+\omega_\mathrm{0}\right)$ (where $\omega_\mathrm{0}$ is the effective EoS parameter during the reheating stage) and thus Eq.~(\ref{T-4}) leads to the following constraint on $\delta$ to satisfy $T_\mathrm{h}\dot{S}_\mathrm{T} + T_\mathrm{m}\dot{S}_\mathrm{m} > 0$:
\begin{eqnarray}
 0 < \delta < \frac{5}{4}~~,
 \label{T-11}
\end{eqnarray}
as the EoS parameter may vary within $\omega_\mathrm{0} = [0,1]$.

\item {\underline{During radiation era}}: In the radiation era, the change of the matter fields' entropy and the horizon entropy are given by,
\begin{eqnarray}
 \dot{S}_\mathrm{m} \propto \frac{3}{a^3H^2}\left(\epsilon - 1\right)~~,
 \label{T-12a}
\end{eqnarray}
and
\begin{eqnarray}
 \dot{S}_\mathrm{T} = \frac{4\pi}{GH}\left(\frac{\delta}{2-\delta}\right) \left(\frac{\pi}{GH^2}\right)^{\delta - 1}~~,
 \label{T-12d}
\end{eqnarray}
respectively. Clearly $\dot{S}_\mathrm{m} > 0$ as $\epsilon$ during the radiation era is larger than unity, and moreover, the positivity of $\dot{S}_\mathrm{T}$ leads to,
\begin{eqnarray}
 0 < \delta < 2~~.
 \label{T-13}
\end{eqnarray}
\end{enumerate}

Due to the reason that $\delta$ remains constant with the cosmic expansion of the universe, all the above constraints on $\delta$ during different cosmic eras get simultaneously fulfilled if it follows
\begin{eqnarray}
 0 < \delta < \mathrm{Min}~\left[2, \frac{5}{4}, 2\right] = \frac{5}{4}~~.
 \label{T-16}
\end{eqnarray}
Here it may be noted that such range of $\delta$ also covers the case of the Bekenstein-Hawking entropy where $\delta = 1$, i.e., the Bekenstein-Hawking entropy also fulfils the requirement of the second law of horizon thermodynamics.
\end{itemize}

Following the above procedure, one can determine the constraints on entropic parameter(s) for other forms of the horizon entropy. Here we give a list for several forms of $S_\mathrm{h}$ \cite{Odintsov:2024hzu}:
\begin{itemize}
 \item For R\'{e}nyi entropy: $S_\mathrm{h} \equiv S_\mathrm{R} = \frac{1}{\alpha}\ln{\left(1 + \alpha S\right)}$ (with $\alpha$ being the parameter), the constraint on the R\'{e}nyi exponent, from inflation to radiation dominated era followed by a reheating stage, comes as:
\begin{eqnarray}
 \alpha > \frac{GH^2_\mathrm{I}}{\pi}~~,
 \label{R-14}
\end{eqnarray}
where $H_\mathrm{I}$ is the Hubble scale during inflation.

\item For the Kaniadakis entropy: $S_\mathrm{h} \equiv S_\mathrm{K} = \frac{1}{K}\mathrm{sinh}(KS)$, the second law of horizon thermodynamics is fulfilled from inflation$\rightarrow$ reheating$\rightarrow$ radiation era if the Kaniadakis exponent obeys the following constraint:
\begin{eqnarray}
 -1.4\left(\frac{GH^2_\mathrm{I}}{\pi}\right) \lesssim K \lesssim 1.4\left(\frac{GH^2_\mathrm{I}}{\pi}\right)~~.
 \label{K-10}
\end{eqnarray}

\item The 4 parameter generalised entropy, given by
\begin{eqnarray}
 S_\mathrm{h} \equiv S_\mathrm{g}\left[\alpha_+,\alpha_-,\beta,\gamma \right] = \frac{1}{\gamma}\left[\left(1 + \frac{\alpha_+}{\beta}~S\right)^{\beta}
 - \left(1 + \frac{\alpha_-}{\beta}~S\right)^{-\beta}\right]~~,
 \label{gen-0}
\end{eqnarray}
is the minimal version of generalised entropy that is able to generalise all the known entropies proposed so far. The parameters should lie within the following constraints in order to validate the second law of horizon thermodynamics:
\begin{eqnarray}
 \frac{\alpha_{\pm}}{\beta} > \frac{GH^2_\mathrm{I}}{\pi}~~~~,~~~~~0 < \beta < \frac{5}{4} ~~~~~~\mathrm{and}~~~~~~~\gamma > 0~~.
 \label{gen-18}
\end{eqnarray}
\end{itemize}

Importantly the above ranges provide $model~independent~constraints$ on entropic parameters (for different entropy functions of apparent horizon) directly from the second law of horizon thermodynamics during a wide range of cosmic era of the universe.

\section{generalised entropy functions}
As mentioned in Eq.~(\ref{HE-4}) and Eq.~(\ref{HE-5}) that different forms of horizon entropy ($S_\mathrm{h}$) leads to different cosmological scenario. In this regard, several entropies have been proposed, like the Tsallis entropy \cite{Tsallis:1987eu}, the R\'{e}nyi entropy \cite{Renyi}, the Barrow entropy \cite{Barrow:2020tzx}, the Sharma-Mittal entropy \cite{SayahianJahromi:2018irq}, the Kaniadakis entropy \cite{Kaniadakis:2005zk}, the Loop Quantum Gravity entropy \cite{Liu:2021dvj} etc. However irrespective of the form, $S_\mathrm{h}$ shares some common properties like:
\begin{itemize}
 \item $S_\mathrm{h}$ is a monotonic increasing function of the Bekenstein-Hawking entropy variable $S = A/(4G)$ (where $A = 4\pi R_\mathrm{h}^2$ denotes the area of the apparent horizon),

 \item $S_\mathrm{h}$ goes to zero in the limit of $S \rightarrow 0$, which can be thought as equivalent of the third law of thermodynamics.
\end{itemize}
Such common properties indicate that there should exist some generalised form of entropy (having few parameters) which can generalise all these known entropies proposed so far at suitable representatives of the entropic parameters. Motivated by this idea, few parameter-dependent generalised entropy functions (both singular and on-singular) have been proposed, which are able to generalise the known entropies like the Tsallis entropy, the R\'{e}nyi entropy, the Barrow entropy, the Sharma-Mittal entropy, the Kaniadakis entropy, the Loop Quantum Gravity entropy. Initially a 6-parameter and a 3-parameter generalised entropy of the forms:
 \begin{eqnarray}
 S_\mathrm{6}(\alpha_\pm,\beta_\pm,\gamma_\pm) = \frac{1}{\alpha_++\alpha_-}\left[\left(1+\frac{\alpha_+}{\beta_+}S^{\gamma_+}\right)^{\beta_+}-\left(1+\frac{\alpha_-}{\beta_-}S^{\gamma_-}\right)^{-\beta_-}\right] \ ,
 \label{intro-1}
\end{eqnarray}
and
\begin{eqnarray}
 S_\mathrm{3}(\alpha,\beta,\gamma) = \frac{1}{\gamma}\left[\left(1+\frac{\alpha}{\beta}S\right)^\beta - 1\right]\ ,
 \label{intro-3}
\end{eqnarray}
were proposed in \cite{Nojiri:2022aof}, where the respective entropic parameters are shown in the bracket. However thereafter this proposal, it was soon realized that the minimum number of parameters required in a generalised entropy function that can generalise all the aforementioned entropies is equal to four. Consequently, the 4-parameter generalised is given by,
\begin{eqnarray}
 S_\mathrm{4}(\alpha_\pm,\beta,\gamma) = \frac{1}{\gamma}\left[\left(1+\frac{\alpha_+}{\beta}S\right)^\beta-\left(1+\frac{\alpha_-}{\beta}S\right)^{-\beta}\right] \ ,
 \label{intro-2}
\end{eqnarray}
where $\left\{\alpha_{\pm},\beta,\gamma\right\}$ are the parameters which are considered to be positive in order to make $S_\mathrm{4}$ as a monotonic increasing function with respect to $S$.

All the above entropies $\left\{S_\mathrm{6},S_\mathrm{4},S_\mathrm{3}\right\}$ possesses a singularity in a different type of cosmological scenario, particularly in bouncing context, as the Bekenstein-Hawking entropy itself diverges in bouncing scenario (at the instant of bounce). Such diverging behaviour is common to all the known entropies (like the Tsallis entropy, the R\'{e}nyi entropy, the Barrow entropy, the Sharma-Mittal entropy, the Kaniadakis entropy and the Loop Quantum gravity entropy). To resolve this issue, a singular-free generalised entropy containing 5-parameters of the form,
\begin{eqnarray}
 S_\mathrm{5}(\alpha_\pm,\beta,\gamma,\epsilon) = \frac{1}{\gamma}\left[\left\{1+\frac{1}{\epsilon}\tanh{\left(\frac{\epsilon\alpha_+}{\beta}S\right)}\right\}^\beta-\left\{1+\frac{1}{\epsilon}\tanh{\left(\frac{\epsilon\alpha_-}{\beta}S\right)}\right\}^{-\beta}\right]\ ,
 \label{intro-4}
\end{eqnarray}
was proposed in \cite{Odintsov:2022qnn}, which turns out to be singular free due to the presence of hyperbolic function and is able to generalise all the entropies known so far. The minimum parameters required for a singular free entropy that is also able to generalize all the known entropies is equal to five. Therefore the minimal constructions of generalised version of entropy is given by the 4-parameter \cite{Nojiri:2022dkr} and the 5-parameter \cite{Odintsov:2022qnn} generalised entropy --- based on universe's evolution, in particular, whether the universe passes through a non-singular bounce (or not) during its cosmic evolution respectively. Various representatives of $\left\{S_\mathrm{6},S_\mathrm{4},S_\mathrm{3},S_\mathrm{5}\right\}$ and their convergence to the known entropies are schematically shown in Table~[\ref{table}]. The wide applications of the generalised entropies towards cosmology as well as towards black holes are addressed in \cite{Nojiri:2022dkr,Odintsov:2022qnn,Odintsov:2023vpj,Bolotin:2023wiw,Lymperis:2023prf,Volovik:2024eni,Odintsov:2024sbo,Brevik:2024nzf}.

\setlength{\tabcolsep}{9pt} 
\renewcommand{\arraystretch}{1.7} 
\begin{table}
\begin{tabular}{ |c|cc| }
\hline
 \multirow{5}{*}{\Large$S_\mathrm{3}$}
& $\gamma=\alpha$ &$S_\mathrm{SM}$  \\
    & $\alpha\rightarrow\infty$ & $S_\mathrm{T},S_\mathrm{B}$ \\
    & $\alpha,\beta\rightarrow 0$ with $\frac{\alpha}{\beta}$ finite & $S_\mathrm{R}$\\
    & $\beta \rightarrow \infty, \gamma=\alpha$ & $S_\mathrm{q}$  \\
    &  &  \\\hline
 \multirow{5}{*}{\Large$S_\mathrm{5}$}
& $\epsilon,\alpha_-\rightarrow0,\alpha_+=\gamma$ &$S_\mathrm{SM}$  \\
    & $\epsilon\rightarrow0,\alpha_-=0,\alpha_+\rightarrow\infty,\gamma=\left(\frac{\alpha_+}{\beta}\right)^\beta$ & $S_\mathrm{T},S_\mathrm{B}$ \\
    & $\epsilon,\beta\rightarrow0,\alpha_-=0,\alpha_+=\gamma$ with $\frac{\alpha_+}{\beta}$ finite & $S_\mathrm{R}$\\
    & $\epsilon,\alpha_-\rightarrow0,\beta\rightarrow\infty,\alpha_+=\gamma$ & $S_\mathrm{q}$  \\
    &$\epsilon\rightarrow0,\beta\rightarrow\infty,\alpha_+=\alpha_-$ & $S_\mathrm{K}$  \\\hline
\end{tabular}
\begin{tabular}{ |c|cc| }
\hline
 \multirow{5}{*}{\Large$S_\mathrm{4}$}
& $\alpha_-=0,\alpha_+=\gamma$ &$S_\mathrm{SM}$  \\
    & $\alpha_+\rightarrow\infty,\alpha_-=0$ & $S_\mathrm{T},S_\mathrm{B}$ \\
    & $\alpha_-=0,\alpha_+=\gamma,\beta\rightarrow0$ with $\frac{\alpha_+}{\beta}$ finite & $S_\mathrm{R}$\\
    & $\beta\rightarrow\infty,\alpha_-=0,\alpha_+=\gamma$ & $S_\mathrm{q}$  \\
    & $\beta \rightarrow \infty, \alpha_+=\alpha_-$ & $S_\mathrm{K}$  \\\hline
     \multirow{5}{*}{\Large$S_\mathrm{6}$}
& $\alpha_-=0,\alpha_+=\gamma_+\beta_+$ &$S_\mathrm{SM}$  \\
    & $\alpha_+=\alpha_-\rightarrow0,\gamma_+=\gamma_-$ & $S_\mathrm{T},S_\mathrm{B}$ \\
    & $\alpha_+,\beta_+\rightarrow0,\gamma_+=1$ with $\frac{\alpha_+}{\beta_+}$ finite & $S_\mathrm{R}$\\
    & $\beta_+\rightarrow\infty,\alpha_-=0,\gamma_+=1$ & $S_\mathrm{q}$  \\
    & $\beta_\pm \rightarrow 0, \alpha_+=\alpha_-,\gamma_\pm=1$ & $S_\mathrm{K}$  \\\hline
\end{tabular}

\caption{\label{table}Schematic table to summarize various representatives of the generalised entropies and their convergence to the known entropies. Here, $S_\mathrm{T} = \textrm{Tsallis entropy}$, $S_\mathrm{B} = \textrm{Barrow entropy}$, $S_\mathrm{R} = \textrm{R\'{e}nyi entropy}$, $S_\mathrm{SM} = \textrm{Sharma-Mittal entropy}$, $S_\mathrm{K} = \textrm{Kaniadakis entropy}$ and $S_\mathrm{q} = \textrm{Loop Quantum gravity entropy}$.}
\end{table}

\section{Primordial gravitational waves (GWs) in entropic cosmology}\label{sec-PGWs}

In this section, we will discuss primordial GWs generated during inflation in the context of entropic cosmology when the entropy of the apparent horizon is given by the 4-parameter generalised entropy ($S_\mathrm{g}$) that has been recently proposed in \cite{Nojiri:2022dkr}. In particular,
\begin{eqnarray}
 S_\mathrm{g}\left[\alpha_+,\alpha_-,\beta,\gamma \right] = \frac{1}{\gamma}\left[\left(1 + \frac{\alpha_+}{\beta}~S\right)^{\beta}
 - \left(1 + \frac{\alpha_-}{\beta}~S\right)^{-\beta}\right]~~,
 \label{PGW-1}
\end{eqnarray}
where $\alpha_{\pm}$, $\beta$ and $\gamma$ are entropic parameters and they are assumed to be positive. During the early stage of the universe, we assume that the matter fields are absent, and then, the FLRW equation corresponding to $S_\mathrm{g}$ results in a constant Hubble parameter (this statement is also true for other forms of horizon entropies). This, in turn, leads to eternal inflation which has no exit mechanism, and moreover, the primordial curvature perturbation is exactly scale-invariant that is inconsistent with the Planck data. Thus in order to have a viable inflation, one may consider that the entropic parameters are not strictly constant, rather they slowly vary with time. One of the possible choices in this regard may be the following \cite{Nojiri:2022dkr}:
\begin{align}
\gamma(N) = \mathrm{exp}\left[\int_{N_\mathrm{f}}^{N}\sigma(N) dN\right]~~~~~~\mathrm{with}~~~~~~~~\sigma(N) = \sigma_0 + \mathrm{e}^{-\left(N_\mathrm{f} - N\right)}~~,
\label{gamma function}
\end{align}
and the other parameters $\alpha_{\pm}$, $\beta$ remain constant. Here $\sigma_0$ is a constant, $N$ denotes the e-folding number with $N_\mathrm{f}$ being the total e-folding number of the inflationary era. With varying $\gamma(N)$, the FLRW equation becomes,
\begin{align}
 -\left(\frac{2\pi}{G}\right)
\left[\frac{\alpha_+\left(1 + \frac{\alpha_+}{\beta}~S\right)^{\beta-1} + \alpha_-\left(1 + \frac{\alpha_-}{\beta}~S\right)^{-\beta-1}}
{\left(1 + \frac{\alpha_+}{\beta}~S\right)^{\beta} - \left(1 + \frac{\alpha_-}{\beta}~S\right)^{-\beta}}\right]\frac{1}{H^3}\frac{dH}{dN} = \sigma(N) \,,
\label{FRW-eq-viable-inf-main1}
\end{align}
on solving which for the Hubble parameter $H = H(N)$, we obtain,
\begin{align}
H(N) = 4\pi M_\mathrm{Pl}\sqrt{\frac{\alpha_+}{\beta}}
\left[\frac{2^{1/(2\beta)}\exp{\left[-\frac{1}{2\beta}\int^{N}\sigma(N)dN\right]}}
{\left\{1 + \sqrt{1 + 4\left(\alpha_+/\alpha_-\right)^{\beta}\exp{\left[-2\int^{N}\sigma(N)dN\right]}}\right\}^{1/(2\beta)}}\right] \, ,
\label{solution-viable-inf-2}
\end{align}
with $\int_0^{N}\sigma(N)dN = N\sigma_0 + \mathrm{e}^{-(N_\mathrm{f} - N)} - \mathrm{e}^{-N_\mathrm{f}}$. The above form of $H(N)$ is compatible with quasi dS inflation with an exit at $N = N_\mathrm{f}$, and moreover, the primordial curvature perturbation and the tensor-to-scalar ratio gets compatible with the Planck data \cite{Akrami:2018odb} for suitable range of the entropic parameters \cite{Nojiri:2022dkr}.

After the inflation ends, the universe enters a reheating phase when the energy density corresponding to $S_\mathrm{g}$ decays to relativistic particles. Here we consider a perturbative reheating scenario which is generally parametrized by a constant EoS parameter ($\omega_\mathrm{eff}$). Therefore the Hubble parameter during the reheating stage is given by,
\begin{eqnarray}
 H(N) = H_\mathrm{f}~\mathrm{exp}\left[-\left(N-N_\mathrm{f}\right)/m\right]~~,
 \label{nreh1}
\end{eqnarray}
where $H_\mathrm{f}$ is the Hubble parameter at the end of inflation (note $H(N)$ is continuous at the junction of $N=N_\mathrm{f}$) and $m$ is the exponent which is related to the reheating EoS parameter by $\omega_\mathrm{eff} = -1+2/(3m)$. The above Hubble parameter during the reheating should be a solution of the main Eq.~(\ref{FRW-eq-viable-inf-main1}), and this is possible for the following form of $\sigma(N)$ \cite{Odintsov:2023vpj}:
\begin{eqnarray}
 \sigma(N) = \left(\frac{2\pi}{G}\right)\frac{e^{2\left(N - N_\mathrm{f}\right)/m}}{mH_\mathrm{f}^2}
 \left[\frac{\alpha_+ \zeta_{+}^{\beta-1} + \alpha_- \zeta_{-}^{-\beta-1}}
{\zeta_{+}^{\beta} - \zeta_{-}^{-\beta}}\right]~,
 \label{nreh2}
\end{eqnarray}
during the reheating stage,where
\begin{eqnarray}
 \zeta_{\pm} = 1 + \frac{\pi\alpha_{\pm}}{\beta GH_\mathrm{f}^2}~e^{2\left(N - N_\mathrm{f}\right)/m}~.
 \nonumber
\end{eqnarray}
Thus as a whole, $\sigma(N)$ has the form of Eq.~(\ref{gamma function}) during inflation and of Eq.~(\ref{nreh2}) during the reheating stage respectively. Consequently, the e-fold for the reheating and the corresponding reheating temperature are given by,
\begin{align}
 N_\mathrm{re}=\frac{2m}{\left(2m-1\right)}\Bigg\{61.6 - \frac{1}{4\beta}\ln{\left[\frac{\beta\mathrm{e}^{-\left(1 + \sigma_0N_\mathrm{f}\right)}\left\{1 + \sqrt{1 + \mathrm{e}^{2\left(1 + \sigma_0N_\mathrm{f}\right)}\left[\left(\frac{1+\sigma_0}{2\beta}\right)^2 - 1\right]}\right\}^2}
{\left(16\pi^2\alpha_+/3\beta\right)^{\beta}~\left\{1+\sigma_0 + 2\beta\right\}}\right]} - N_\mathrm{f}\Bigg\}
 \label{et-10}
\end{align}
and
\begin{eqnarray}
 T_\mathrm{re} = H_\mathrm{i}\left(\frac{43}{11g_\mathrm{re}}\right)^{1/3}\left(\frac{T_0}{k/a_0}\right)e^{-\left(N_\mathrm{f} + N_\mathrm{re}\right)}~~,
 \label{et-4}
\end{eqnarray}
respectively. The reheating phenomenology requires $N_\mathrm{re} > 0$ and $T_\mathrm{re} > T_\mathrm{BBN} \approx 10^{-2}\mathrm{GeV}$. The following ranges of the entropic parameters lead to a viable phenomenology during inflation as well as during the reheating \cite{Odintsov:2023vpj}:
\begin{table}[h]
  \centering
 {%
  \begin{tabular}{|c|c|c|c|c|}
   \hline
    Viable choices of $N_\mathrm{f}$ & Viable range of $\sigma_0$ & Viable range of $\beta$ & Viable range of $\left(\frac{\alpha_+}{\alpha_-}\right)^{\beta}$ & Reheating EoS\\

   \hline
  \hline
   (1) Set-1: $N_\mathrm{f} = 50$ & $\sigma_0 = [0.0127,0.0166]$ & (a) $0.05 < \beta < 0.10$ & $2\times10^{5} < \left(\frac{\alpha_+}{\alpha_-}\right)^{\beta} < 8.5\times10^{5}$ & $\frac{1}{3} < w_\mathrm{eff} < 1$\\
   \hline
     & & (b) $0.10 < \beta < 0.35$ & $7.5 < \left(\frac{\alpha_+}{\alpha_-}\right)^{\beta} < 2\times10^{5}$ & $-\frac{1}{3} < w_\mathrm{eff} < \frac{1}{3}$\\
     \hline
    (2) Set-2: $N_\mathrm{f} = 55$ & $\sigma_0 = [0.0129,0.0166]$ & (a) $0.06 < \beta < 0.22$ & $4\times10^{4} < \left(\frac{\alpha_+}{\alpha_-}\right)^{\beta} < 5\times10^{5}$ & $\frac{1}{3} < w_\mathrm{eff} < 1$\\
   \hline
     & & (b) $0.22 < \beta < 0.40$ & $7.5 < \left(\frac{\alpha_+}{\alpha_-}\right)^{\beta} < 4\times10^{4}$ & $-\frac{1}{3} < w_\mathrm{eff} < \frac{1}{3}$\\
     \hline
     (3) Set-3: $N_\mathrm{f} = 60$ & $\sigma_0 = [0.0130,0.0166]$ & (a) $0.08 < \beta < 0.40$ & $7.5 < \left(\frac{\alpha_+}{\alpha_-}\right)^{\beta} < 3\times10^{5}$ & $\frac{1}{3} < w_\mathrm{eff} < 1$\\
   \hline
     \hline
  \end{tabular}%
 }
  \caption{Viable ranges on entropic parameters coming from both the inflation and reheating phenomenology for three different choices of $N_\mathrm{f}$. Here it is important to mention that $\omega_\mathrm{eff}$ needs to be greater than $1/3$ for $N_\mathrm{f} \gtrsim 57$.}
  \label{Table-0}
 \end{table}

Having set the background evolution, we now address the spectrum of primordial GWs generated during inflation in the context of generalised entropic cosmology \cite{Odintsov:2024sbo}. If $h_{ij}(t,\vec{x})$ be the tensor perturbation characterizing GWs over a spatially flat FLRW spacetime, then the spacetime metric can be expressed by,
\begin{eqnarray}
 ds^2 = -dt^2 + a^2(t)\left[\left(\delta_{ij} + h_{ij}\right)dx^{i}dx^{j}\right]~~.
 \label{I-1}
\end{eqnarray}
On quantizing $h_{ij}(t,\vec{x})$, we may write the mode expansion as,
\begin{eqnarray}
 \hat{h}_{ij}(t,\vec{x}) = \sum_{\lambda=+,\times}\int \frac{d^3\vec{k}}{(2\pi)^{3/2}}\left[\hat{a}_k^{\lambda}\epsilon_{ij}^{\lambda}(\vec{k})h(k,t)\mathrm{e}^{i\vec{k}.\vec{x}} + c.c.\right]~~,
 \label{I-3}
\end{eqnarray}
where $\lambda = +,\times$ denotes two types of polarizations of the GWs, $\epsilon_{ij}^{\lambda}(\vec{k})$ are the polarisation tensor and $\hat{a}_k$ ($\hat{a}_k^{+}$) are the annihilation (creation) operators respectively that satisfy the usual commutation rules. Moreover, from the transverse condition of GWs, i.e., due to $\partial_{i}h^{ij} = 0$, one immediately obtains $k^{i}\epsilon_{ij}^{\lambda}(\vec{k}) = 0$. The Fourier mode  $h(k,t)$ obeys,
\begin{eqnarray}
 \ddot{h}(k,t) + 3H\dot{h}(k,t) + \frac{k^2}{a^2}h(k,t) = 0~~.
 \label{I-4}
\end{eqnarray}
As described above, the background Hubble parameter is almost constant during the inflation, and thus Eq.~(\ref{I-4}) is solved for $h(k,t)$ during the same as follows:
\begin{align}
    h(k,\eta) = -\sqrt{\frac{2}{k}}\left(\frac{H_\mathrm{i}}{M_\mathrm{Pl}}\right)\eta \,  e^{-ik\eta}\left(1-\frac{i}{k\eta}\right) \label{h_inflation} \ ,
\end{align}
where $H_\mathrm{i}$ is the constant Hubble parameter during inflation and can be obtained from Eq.~(\ref{solution-viable-inf-2}) at $N = 0$. For the post-inflationary evolution, let us introduce the transfer function $\chi(k,\eta)$ as
\begin{equation}\label{I-9}
    h(k,\eta)=\left[\lim_{|k\eta|\ll 1}h(k,\eta)\right]\chi(k,\eta) = i\sqrt{\frac{2}{k^3}}\left(\frac{H_\mathrm{i}}{M_\mathrm{Pl}}\right)\chi(k,\eta) \ ,
\end{equation}
in terms of which, Eq.~(\ref{I-4}) takes the following form,
\begin{equation}\label{EoM_chi_general}
    \Ddot{\chi}(k,t)+3H\dot{\chi}(k,t)+\frac{k^2}{a^2}\chi(k,t)=0 \ .
\end{equation}
During the reheating stage, $H \propto A^{-\frac{3}{2}(1+w_\mathrm{eff})}$ (with $A = a/a_\mathrm{f}$ is the rescaled scale factor, note that $A = 1$ at the end of inflation), and consequently, Eq.~(\ref{EoM_chi_general}) is solved as,
\begin{equation}\label{chi_reheating}
    \chi^\mathrm{RH}(k,A)=A^{-\frac{3+3w_\mathrm{eff}}{4}}\left[C(k)~J_\nu\left(\frac{2k/k_\mathrm{re}}{1+3w_\mathrm{eff}}\left(\frac{A}{A_\mathrm{re}}\right)^\frac{1+3w_\mathrm{eff}}{2}\right) + D(k)~J_{-\nu}\left(\frac{2k/k_\mathrm{re}}{1+3w_\mathrm{eff}}\left(\frac{A}{A_\mathrm{re}}\right)^\frac{1+3w_\mathrm{eff}}{2}\right) \right] \ ,
\end{equation}
where $\nu = (1-3m)/(2+2m)$ and $k_\mathrm{re}$ represent the mode that re-enters the horizon at the end of reheating. Moreover, $C(k)$ and $D(k)$ are the integration constants which can be determined from the continuity condition of the transfer function at the junction between inflation to reheating, given by,
\begin{equation}\label{Reh-3}
    \chi(k,A=1) = 1 \hspace{0.5cm} \text{and} \hspace{0.5cm}\frac{d\chi}{dA}\bigg|_{A=1}=0
\end{equation}
respectively. Furthermore, the transfer function during radiation follows from Eq.~(\ref{I-4}) by using $H \propto A^{-2}$, and is given by,
\begin{eqnarray}\label{rd-3}
    \chi^\mathrm{RD}(k,A)&=&\frac{e^{-ib(A-A_\mathrm{re})}}{2A}\left[\left(A_\mathrm{re} - \frac{1}{ib}\right)\chi^\mathrm{RH}(k,A_\mathrm{re}) - \left(\frac{A_\mathrm{re}}{ib}\right)\frac{d\chi^\mathrm{RH}(k,A_\mathrm{re})}{dA}\right]\nonumber\\
    &+&\frac{e^{ib(A-A_\mathrm{re})}}{2A}\left[\left(A_\mathrm{re} + \frac{1}{ib}\right)\chi^\mathrm{RH}(k,A_\mathrm{re}) + \left(\frac{A_\mathrm{re}}{ib}\right)\frac{d\chi^\mathrm{RH}(k,A_\mathrm{re})}{dA}\right]  \ .
\end{eqnarray}
Here $\chi^\mathrm{RH}(k, A_\mathrm{re})$ represents the transfer function at the end of reheating, and thus $\chi(k, A)$ becomes continuous at the junction between reheating to radiation occurring at $A = A_\mathrm{re}$.

The dimensionless energy density parameter $\Omega_\mathrm{GW}^\mathrm{(0)}(k)$ today (i.e., at present epoch) is given by,
\begin{eqnarray}
 \Omega_\mathrm{GW}^\mathrm{(0)}(k)h^2 = \frac{1}{6\pi^2}\left(\frac{g_{r,0}}{g_{r,eq}}\right)^{1/3}\Omega_\mathrm{R}h^2\left(\frac{H_\mathrm{i}}{M_\mathrm{Pl}}\right)^2\left\{A^2\left|\frac{d\chi^\mathrm{RD}(k,A)}{dA}\right|^2 + b^2A^2\left|\chi^\mathrm{RD}(k,A)\right|^2\right\}~~,
 \label{rd-8}
\end{eqnarray}
where $\Omega_\mathrm{R}$ denotes the present-day dimensionless energy density of radiation, $g_{r,eq}$ and $g_{r,0}$ represent the number of relativistic degrees of freedom at matter-radiation equality and today respectively. The above solutions of $\chi(k,A)$ lead to the following forms of $\Omega_\mathrm{GW}^\mathrm{(0)}(k)$ for different modes.
\begin{itemize}
 \item For the modes that re-enter the horizon during the radiation era, i.e., for $k < k_\mathrm{re}$ (where $k_\mathrm{re}$ represents the mode which re-enters the horizon at the end of reheating):
 \begin{equation}\label{CI-3}
    \boxed{\ \Omega_\mathrm{GW}^\mathrm{(0)}(k)h^2 \simeq \left(\frac{1}{6\pi^2}\right)\Omega_\mathrm{R}h^2\left(\frac{H_\mathrm{i}}{M_\mathrm{Pl}}\right)^2 \ , \ }
\end{equation}
where we have assumed that $g_{r,0} = g_{r,eq}$.

\item For the modes that re-enter the horizon during the reheating stage, i.e., for $k_\mathrm{re} < k < k_\mathrm{f}$ (where $k_\mathrm{f}$ is the mode that re-enters the horizon at the end of inflation):
\begin{equation}\label{CII-5}
    \boxed{\ \Omega_\mathrm{GW}^\mathrm{(0)}(k)h^2 \simeq \left(\frac{1}{6\pi^2}\right)\Omega_\mathrm{R}h^2\left(\frac{H_\mathrm{i}}{M_\mathrm{Pl}}\right)^2 \big(1+3w_\mathrm{eff}\big)^\frac{4}{1+3w_\mathrm{eff}}\left(\frac{\Gamma(1-\nu)}{\sqrt{\pi}}\right)^2\left(\frac{k}{k_\mathrm{re}}\right)^{2\left(\frac{3w_\mathrm{eff}-1}{3w_\mathrm{eff}+1}\right)} \ , \ }
\end{equation}
where, once again, we have assumed that $g_{r,0} = g_{r,eq}$.
\end{itemize}

\begin{figure}[!h]
\begin{center}
\centering
\includegraphics[width=3.2in,height=2.2in]{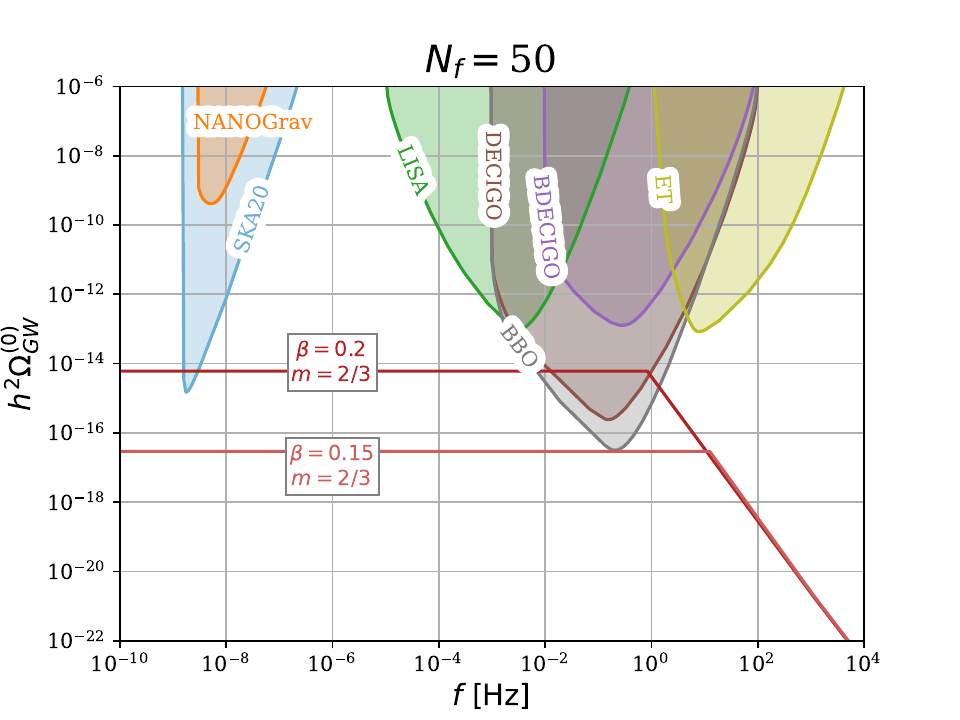}
\includegraphics[width=3.2in,height=2.2in]{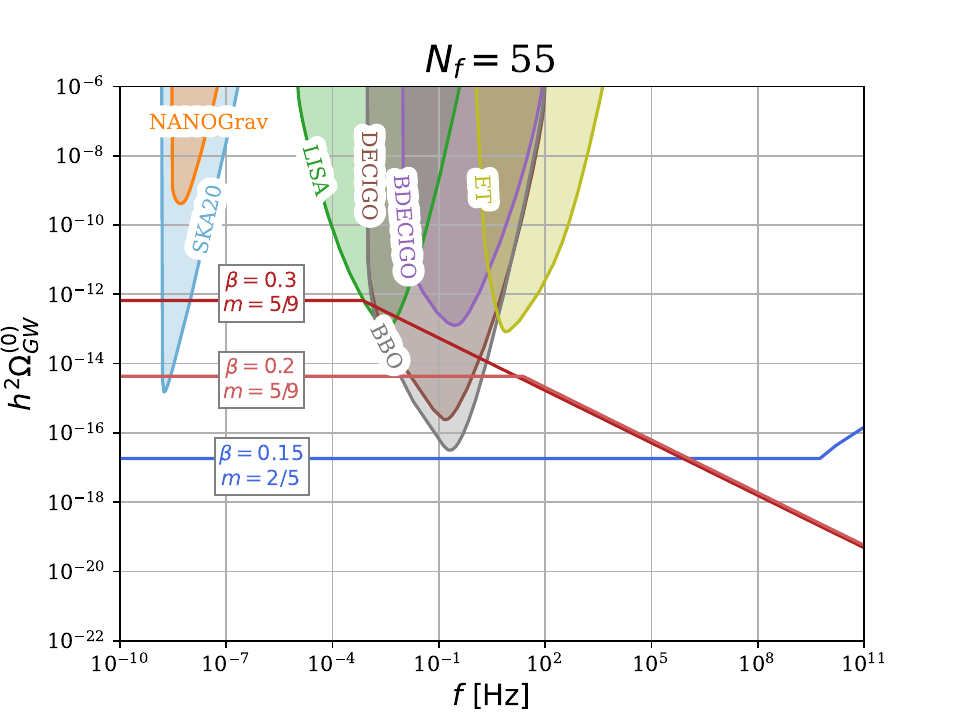}
\caption{{\color{blue}Left Plot}: $\Omega_\mathrm{GW}^\mathrm{(0)}$ vs. $\mathrm{f}$[Hz] for $N_\mathrm{f} = 50$; {\color{blue}Right Plot}: $\Omega_\mathrm{GW}^\mathrm{(0)}$ vs. $\mathrm{f}$[Hz] for $N_\mathrm{f} = 55$. In both the plots, we consider a set of values of the entropic parameters $\beta$ and $m$, and moreover, the other entropic parameters, namely $\sigma_0$ and $\alpha_{+}$, are taken as $\sigma_0= 0.015$ and $\alpha_{+}/\beta = 10^{-6}$, see Table.~[\ref{Table-0}]. Clearly, the GWs spectra are flat for $k<k_\mathrm{re}$, while it has a non-zero tilt in the domain $k_\mathrm{re} < k < k_\mathrm{f}$. In particular, we take $m = 2/3$, $5/9$ and $2/5$ which lead to the indices $n_\mathrm{GW} = -2$, $-1/2$ and $2/3$ respectively.}
\label{plot-R1}
\end{center}
\end{figure}

As a whole, Eq.~(\ref{CI-3}) and Eq.~(\ref{CII-5}) clearly demonstrate that the GWs today has a flat spectrum for the modes which re-enter the horizon during the radiation dominated era; while the spectrum is tilted over the modes that re-enter the horizon during the reheating era. The amount of such tilt is given by,
\begin{eqnarray}
 n_\mathrm{GW} = 2\left(\frac{3w_\mathrm{eff}-1}{3w_\mathrm{eff}+1}\right) = \frac{2(1-2m)}{1-m}~,
\end{eqnarray}
which is blue for $m < \frac{1}{2}$, and red for $m > \frac{1}{2}$. In Fig.~[\ref{plot-R1}], the GWs spectrum today are plotted for a set of values of the entropic parameters $\beta$, $\sigma_0$ and $m$, and moreover, for two different inflationary e-folding number: $N_\mathrm{f} = 50$ and $55$ respectively. The set of values of the entropic parameters are consistent with their viable ranges from Table.~[\ref{Table-0}]. The figures clearly illustrate the qualitative features as discussed, and therefore, if future observatories can detect the signal of primordial GWs, then our theoretical expectation carried in the present work may provide a possible tool for the measurement of the generalised entropic parameters.

\section{Non-singular bounce in entropic cosmology}

In order to have a bounce in entropic cosmology, the entropy of the apparent horizon itself needs to be non-singular at the instance of the vanishing Hubble parameter ($H = 0$). This is unlike to the 4-parameter generalised entropy (or even to the other known entropies proposed so far like the Tsallis entropy, Renyi entropy etc.) which becomes singular at $H = 0$. With the spirit of addressing a non-singular bounce, a new singular-free entropy function has been recently proposed in \cite{Odintsov:2022qnn}, and is given by,
\begin{eqnarray}
S_\mathrm{ns}\left[\alpha_{\pm},\beta,\gamma,\epsilon\right] = \frac{1}{\gamma}\bigg[\left\{1 + \frac{1}{\epsilon}\tanh\left(\frac{\epsilon \alpha_+}{\beta}~S\right)\right\}^{\beta}
- \left\{1 + \frac{1}{\epsilon}\tanh\left(\frac{\epsilon \alpha_-}{\beta}~S\right)\right\}^{-\beta}\bigg]~~,
\label{gen-entropy-b}
\end{eqnarray}
where $\alpha_{\pm}$, $\beta$, $\gamma$ and $\epsilon$ are the parameters which are considered to be positive, $S =\pi/(GH^2)$ symbolizes
the Bekenstein-Hawking entropy and the suffix `ns' stands for `non-singular'. Below we will use the following notation:
\begin{eqnarray}
 \chi_{\pm} = 1 + \frac{1}{\epsilon}\tanh\left(\frac{\epsilon \alpha_{\pm}}{\beta}~S\right)~.
 \nonumber
\end{eqnarray}
Note, due to the tan hyperbolic nature, the above form of entropy remains finite at the instance of $H = 0$, (i.e., at the time of bounce). However $S_\mathrm{ns}$, with constant parameters, does not provide viable cosmology, and thus we consider the parameter $\gamma$ to vary with time, and all the other parameters remain fixed, i.e.,
\begin{eqnarray}
 \gamma = \gamma(N)~~,
 \label{gamma}
\end{eqnarray}
with $N$ being the e-fold number of the universe. In effect of $\gamma = \gamma(N)$, the modified Friedmann equations corresponding to the $S_\mathrm{ns}$ is given by:
\begin{eqnarray}
\left[\frac{\alpha_{+}~\mathrm{sech}^2\left(\frac{\epsilon \alpha_+}{\beta}S\right)
\chi_{+}^{\beta-1}
+ \alpha_{-}~\mathrm{sech}^2\left(\frac{\epsilon\alpha_-}{\beta}S\right)\chi_{-}^{-\beta-1}}
{\chi_{+}^{\beta} -
\chi_{-}^{-\beta}}\right]dS = \frac{\gamma'(N)}{\gamma(N)}dN
\nonumber
\end{eqnarray}
where an overprime denotes $\frac{d}{d\eta}$. Integrating the above equation, one obtains,
\begin{eqnarray}
 \tanh{\left(\frac{\epsilon \pi\alpha}{\beta GH^2}\right)} = \left\{\frac{\gamma(N) + \sqrt{\gamma^2(N) + 4}}{2}\right\}^{1/\beta} - 1~~.
 \label{bounce-6}
\end{eqnarray}
where we take $\alpha_+ = \alpha_- = \alpha$ (say, without losing any generality) in order to extract an explicit solution of $H(N)$ which indeed depends on the explicit form of $\gamma(N)$. In the following we consider two symmetric bounce cases, particularly the exponential bounce and the quasi-matter bounce cases, and will determine the associated form of $\gamma(N)$ by using Eq.(\ref{bounce-6}).
\begin{enumerate}
 \item The scale factor,
 \begin{eqnarray}
 a(t) = \mathrm{exp}\left(a_0t^2\right)~~,
 \label{exp bounce-1}
\end{eqnarray}
describes the exponential bounce where the bounce happens at $t = 0$. Here $a_0$ is a constant having mass dimension [+2], which is actually related with the entropic parameters of $S_\mathrm{ns}$ and thus, without losing any generality, we take $a_0 = \frac{\epsilon \pi\alpha}{4G\beta}$. Such an exponential bounce can be achieved from singular free entropic cosmology provided the $\gamma(N)$ is given by,
\begin{eqnarray}
 \gamma(N) = \left\{1 + \frac{1}{\epsilon}\tanh\left(\frac{1}{N}\right)\right\}^{\beta} -
\left\{1 + \frac{1}{\epsilon}\tanh\left(\frac{1}{N}\right)\right\}^{-\beta}~~.
\label{exp bounce-4}
\end{eqnarray}
\item The quasi-matter bounce is described by the scale factor:
\begin{eqnarray}
 a(t) = \left[1 + a_0\left(\frac{t}{t_0}\right)^2\right]^n~~,
 \label{matter bounce-1}
\end{eqnarray}
where $n$, $a_0$ and $t_0$ are connected to the entropic parameters. In particular, one may take: $n = \sqrt{\alpha}$, $a_0 = \frac{\pi}{4\beta}$ and $t_0 = \sqrt{G/\epsilon}$, with $G$ being the gravitational constant. Consequently, the $\gamma(N)$ which leads to such quasi-matter bounce, comes as,
\begin{eqnarray}
\gamma(N) = \left\{1 + \frac{1}{\epsilon}\tanh\left[\mathrm{e}^{-N/\sqrt{\alpha}}\left(\mathrm{e}^{N/\sqrt{\alpha}} - 1\right)^{\frac{1}{2}}\right]\right\}^{\beta} -
\left\{1 + \frac{1}{\epsilon}\tanh\left[\mathrm{e}^{-N/\sqrt{\alpha}}\left(\mathrm{e}^{N/\sqrt{\alpha}} - 1\right)^{\frac{1}{2}}\right]\right\}^{-\beta}~~.
\label{matter bounce-6}
\end{eqnarray}
\end{enumerate}
At this stage it deserves mentioning that the comoving Hubble radius in the case of exponential bounce monotonically decreases with time and asymptotically goes to zero at both sides of the bounce. This results in the fact that the primordial perturbation modes generate near the bounce where all the modes lie in the sub-Hubble regime, and moreover, the perturbation modes in the distant past remain outside of the Hubble radius. As a result, the exponential bounce suffers from a horizon problem. This is unlike the quasi-matter bounce where the comoving Hubble radius monotonically increases with the cosmic time and eventually diverges at the asymptotic regime. Thus the perturbation modes generate and lie in the sub-Hubble regime in the distant past far away from the bounce -- this resolves the horizon problem in this case. Moreover, the quasi-matter bounce also leads to viable observable quantities consistent with the Planck data \cite{Odintsov:2022qnn}.

\section{Brief discussion on future perspectives}

In Sec.~[\ref{sec-law-1-MGT}], we have discussed how a gravity theory is linked with a specific form of horizon entropy through the first law of horizon thermodynamics. Therefore it is important to understand what will be the gravity theory corresponding to the 4-parameter generalised entropy. This is important because the 4-parameter generalised entropy is able to generalise all the known entropies, thus the equivalent gravity theory to such generalised entropy must have rich consequences in the context of cosmology as well as in black hole physics. It is evident from Sec.~[\ref{sec-PGWs}] that the theoretical expectation of GWs spectra, based on the 4-parameter generalised entropy, does not intersect the sensitivity curve of the NANOGrav. This may indicate that the standard inflationary evolution may not be the full story of the early universe. Thus a modified inflationary evolution, for instance, a short deceleration epoch inside inflation, may be required to corroborate the theoretical GWs spectra with the NANOGrav data. The aspect of the generalised entropy in such modified cosmic evolution and its consistency with the NANOGrav data has significance in its own right. Apart from the early universe scenario, there are few unsolvable issues of entropic cosmology yet in the context of the dark energy era. For instance -- (a) can we obtain a viable reason from entropic cosmology regarding the cosmic transition of the universe from the standard deceleration to the late time acceleration? (b) what is the view of entropic cosmology on the Hubble tension issue as well as on the LCDM epoch? etc. Moreover with development of entropic cosmology, it remains to study black holes and compact objects which are obtained with entropic modification of FLRW equations. These issues are timely and perhaps will be studied in future.

\section*{``Data Availability Statement'' and ``Conflicts of Interest''}
This review article is theoretical and no data has been used. Moreover there is no ``Conflicts of Interest''.

\end{document}